\documentclass[useAMS,usenatbib]{mn2e}
\usepackage{amssymb}
\usepackage{graphicx}

\title[]{Asteroseismology of DAV star HS0507+0434B, including the core composition profiles}
\author[Y. H. Chen, Y. Li]{Y. H. Chen$^{1,2,3}$\thanks{E-mail: yanhuichen1987@ynao.ac.cn, ly@ynao.ac.cn} and Y. Li$^{1,2}$\\
$^{1}$Yunnan Astronomical Observatory, Chinese Academy of Sciences, Kunming 650011, China\\
$^{2}$Key Laboratory for the Structure and Evolution of Celestial Objects, Chinese Academy of Sciences, Kunming 650011, China\\
$^{3}$University of Chinese Academy of Sciences, Beijing 100049, China}

\begin{document}

\date{Accepted: }

\pagerange{\pageref{firstpage}--\pageref{lastpage}} \pubyear{????}

\maketitle

\label{firstpage}

\begin{abstract}
The DAV star, HS0507+0434B, was observed by Fu et al. (2013) in 2007 and from 2009 December to 2010 January. There were a total of six triplets with nearly equal split, which were identified as $l$ = 1 modes caused by rotation. In order to fit the six $l$ = 1 modes, grids of white dwarf models are generated by WDEC. For the core composition profiles, we choose the linear fittings to carbon profile of white dwarf models from MESA, which can be considered as results of real nuclear burning process. Coupled with diffusion equilibrium H/He and He/C mixtures, we make grids of models in WDEC and do asteroseismology works on HS0507+0434B. There is a total of 9.50 seconds error for our best-fitting model, which is smaller than the result (a total of 22.1 seconds error) of Fu et al. (2013), when fitting the six $l$ = 1 modes. The two other previous identified pulsation mode frequencies (286.1 s and 743.40 s) may also be well fitted by our best-fitting model. The model parameters are $T_{eff}$ = 11450 K, log$g$ = 8.088, $M_{*}$ = 0.640 $M_{\odot}$, log($M_{H}$/$M_{*}$) = -6, and log($M_{He}$/$M_{*}$) = -3.
\end{abstract}


\section{Introduction}

About 98\% of all stars will evolve to be white dwarfs (Winget \& Kepler 2008). Therefore, the study of white dwarfs is of a universal significance. A white dwarf is composed of a dense degenerate core and a thin layer of ideal gas atmospheres. Those white dwarfs, classified as DO and DB, have helium-dominated atmospheres. For DA white dwarfs, there are hydrogen-dominated atmospheres on the surface. DA white dwarfs comprise roughly 80\% of all white dwarfs (Bischoff-Kim \& Metcalfe 2011). The thermonuclear burning is basically ceased and therefore they are cooling down by shrinking and eradiating. On the cooling curve, there are DOV, DBV, and DAV pulsation instability strips. The DAV stars are non-radial $g$-mode pulsators. The $g$-modes, with buoyancy acting as restoring force, can pulsate basically from the inner core to the outer regions of its atmosphere except for the convection zones. The pulsation mode can be characterized by three indices ($k$, $l$, $m$), where, $k$ is the radial order, $l$ the spherical degree, and $m$ the azimuthal number. According to the asymptotic theory for $g$-modes (Tassoul 1980), the frequencies of low-degree and high-order $g$-modes are given by
\begin{equation}
\omega=\frac{\sqrt{l(l+1)}\int_{0}^{R}N\frac{dr}{r}}{\pi(k+l/2+\alpha_{g})}.
\end{equation}
\noindent In Eq. (1), $N$ is the buoyancy frequency, $R$ the stellar radius, $\omega$ the pulsation frequency, and $\alpha_{g}$ a phase constant.

Asteroseismology is a powerful tool to detect the inner structure of white dwarfs, such as PG1159-035 (e.g. Kawaler \& Bradley 1994; Costa et al. 2008; Costa \& Kepler 2008) and GD358 (e.g. Kepler et al. 2003; Provencal et al. 2009). For DAV stars, there are usually only a few modes observed because of the small amplitudes. Therefore, the construction of physical and realistic DAV models is especially important. With spectroscopic observations, Fontaine et al. (2003) tried to inverse the total stellar mass of 12 DAV stars adopting full carbon core composition (Wood 1995). It means that the core is composed of only carbon without any oxygen. In fact, in the reaction of helium burning, a part of the product ${}^{12}{\rm C}$ will continue to capture ${}^{4}{\rm He}$ to generate ${}^{16}{\rm O}$. Castanheira \& Kepler (2008; 2009) did the asteroseismology study for 83 DAV stars adopting the homogeneous core composition. It means that there are flat profiles for carbon and oxygen namely C/O = 50/50. Fu et al. (2013) did the asteroseismology study for HS0507+0434B, also adopting the homogeneous core composition (Dolez \& Vauclair 1981). Studying the theoretical models for DA white dwarfs, Bradley (1996) made the core of 20\% carbon (80\% oxygen) out to 0.75 $M_{*}$ and then linear changing to pure carbon by 0.90 $M_{*}$, where $M_{*}$ is the total stellar mass. All of the three core compositions are rough approximations, rather than results of thermonuclear burning process. Romero et al. (2012) made fully evolutionary models to do the asteroseismology works on 44 bright DAV stars. The white dwarf models were evolved from the main-sequence stars by $LPCODE$ evolutionary code. Inspired by the evolution method, we try to do a sequence evolution by Modules for Experiments in Stellar Astrophysics (MESA). MESA is powerful enough to evolve a star, such as one solar mass, from pre-main sequence to white dwarf stage (Paxton et al. 2011). The core compositions are results of nuclear burning process.

For asteroseismology, long-time photometric observations are necessary. Using appropriate analysis methods, such as Fourier analysis and multi-frequency sine-wave fitting method, the pulsation mode frequencies can be found basically. According to the rotational split, the spherical degree can be identified. The modes will be identified as $l$ = 1 ones if there are triplets observed, such as some modes of the DAV star EC14012-1446 (Handler et al. 2008; Provencal et al. 2012). If there are quintuplets observed, the modes will be identified as $l$ = 2 ones. However, as far as we known, there are no quintuplets observed for DAV stars. Usually, the observed modes have no frequency split. The $l$ identification of these modes is difficult. They are likely to be $l$ = 1 modes, also may be $l$ = 2 modes, even $l$ = 3 or 4 modes (Thompson et al. 2008). The effect of spherical harmonic degree on the asteroseismology work on DAV stars had been discussed by Chen \& Li (2013). With enough identified pulsation mode frequencies, we have the chance to do the asteroseismology study. HS0507+0434B was observed by Fu et al. (2013) in 2007 and from 2009 December to 2010 January. There are a total of six triplets with nearly equal split, which can be identified as $l$ = 1 modes reliably by rotation. In the perspective of mode identification, it is suitable for HS0507+0434B to do the asteroseismology work. With no uncertainties in mode identification, the asteroseismology work is more reliable.

In this paper, we try to study the core composition of HS0507+0434B by asteroseismology method. In Sect. 2, we review the previous works on HS0507+0434B. Then, we show our input physics and model calculations on HS0507+0434B in Sect. 3, including the detailed method to study the core composition. In Sect. 4, we display our asteroseismology results and compare the results to the previous works. Then, we obtain our asteroseismology parameters for HS0507+0434B, including the core composition profiles. At last, we do some discussions and summarize our conclusions in Sect. 5.

\section{The previous works on HS0507+0434B}

\begin{table}
\begin{center}
\begin{tabular}{lll}
\hline
$P_{handler}(s)$&$P_{kotak}(s)$ \\
\hline
743.373 &  743.0  \\
557.628 &  557.7  \\
556.542 &         \\
555.341 &         \\
446.136 &  446.2  \\
445.309 &         \\
444.641 &  444.8  \\
355.826 &  355.8  \\
355.366 &         \\
354.875 &  354.9  \\
        &  286.1  \\
\hline
\end{tabular}
\caption{Observed modes from Handler et al. (2002) and Kotak et al. (2002).}
\end{center}
\end{table}

Jordan et al. (1998) discovered HS0507+0434B as a ZZ Ceti variable by the Hamburg Quasar Survey. Handler et al. (2002) made a week of single site observations. They found one singlet and three triplets nearly equal split. Kotak et al. (2002) also made a detailed analysis of time-resolved optical spectra of HS0507+0434B and found seven real modes. Their results are shown in Table 1. Taking the high amplitude modes shown in Handler et al. (2002), Romero et al. (2012) made the asteroseismology study for HS0507+0434B. Their grids of models are fully evolutionary white dwarfs, as mentioned above. The observed modes adopted and the best-fitting model results are shown in Table 2. The commonly used parameter $\phi$ is introduced, which is expressed by,

\begin{equation}
\phi=\frac{1}{n}\sum(|P_{mod}-P_{obs}|).
\end{equation}
\noindent In Eq. (2), $n$ is the number of the observed modes we adopt, $P_{mod}$ the model period, and $P_{obs}$ the period observed. The model of the smallest $\Phi$ is considered as the best-fitting one. Romero et al. (2012) found their best-fitting model with $T_{eff}$ = 12257 $\pm$ 135 K, log$g$ = 8.10 $\pm$ 0.06, $M_{*}$ = 0.660 $\pm$ 0.023 $M_{\odot}$, $M_{H}$/$M_{*}$ = (5.68 $\pm$ 1.94)$\times$$10^{-5}$, and $M_{He}$/$M_{*}$ = 1.21$\times$$10^{-2}$. The parameter $\Phi$ is 0.778 seconds, which is very small.

\begin{table}
\begin{center}
\begin{tabular}{llll}
\hline
$P_{obs}(s)$&$P_{mod}(s)$ & $\phi$(s) \\
\hline
743.40  &  742.920 & 0.778\\
555.30  &  556.767 &      \\
446.20  &  446.429 &      \\
355.80  &  356.737 &      \\
\hline
\end{tabular}
\caption{The best-fitting model results of Romero et al. (2012).}
\end{center}
\end{table}

Fu et al. (2013) carried out multi-site observations campaigns for HS0507+0434B in 2007 and from 2009 December to 2010 January. They obtained six triplets and did detailed period to period fittings for the identified six $l$ = 1 modes. Their best-fitting model results are shown in Table 3. The best-fitting model parameters are $T_{eff}$ = 12460 K, $M_{*}$ = 0.675 $M_{\odot}$, log($M_{H}$/$M_{*}$) = -8.5, and log($M_{He}$/$M_{*}$) = -2. The parameter $\Phi$ is 3.683 seconds for their best-fitting model. It is interesting to note that the mode nearly 743.40 s, observed by Handler et al. (2002), Kotak et al. (2002), and fitted by Romero et al. (2012), disappears. Instead, there are modes of 750.3 s, 748.6 s, and 746.1 s being identified as a triplet. If 748.6 s is an $l$ = 1 mode, the mode of 743.40 s is then impossible to be an $l$ = 1 mode. In the works of Romero et al. (2012), if we use the mode of 742.920 s to fit 748. 6 s, the error will be a little large. For the asteroseismology works of Fu et al. (2013), taking the homogeneous mixtures of carbon and oxygen for the degenerate core, $\Phi$ is 3.683 seconds.

\begin{table}
\begin{center}
\begin{tabular}{llll}
\hline
$P_{obs}(s)$&$P_{mod}(s)$ & $\phi$(s) \\
\hline
748.6   &  746.0 & 3.683\\
697.6   &  689.9 &      \\
655.9   &  654.6 &      \\
556.5   &  560.5 &      \\
445.3   &  449.2 &      \\
355.3   &  357.9 &      \\
\hline
\end{tabular}
\caption{The best-fitting model results of Fu et al. (2013).}
\end{center}
\end{table}

\begin{table}
\begin{center}
\begin{tabular}{lccccc}
\hline
                     &Fontaine       &Koester          &Gianninas       \\
\hline
$T_{eff}$(K)         &11630$\pm$200  &11488$\pm$18     &12290$\pm$186   \\
log$g$               &8.17$\pm$0.05  &8.057$\pm$0.008  &8.24$\pm$0.05   \\
$M_{*}$/$M_{\odot}$  &0.71           &                 &0.75$\pm$0.03   \\
\hline
\end{tabular}
\caption{The best-fitting model parameters for the previous spectral works of Fontaine et al. (2003), Koester et al. (2009), and Gianninas et al. (2011).}
\end{center}
\end{table}

From the spectroscopic results, Fontaine et al. (2003) and Bergeron et al. (2004) gave the effective temperature $T_{eff}$ = 11630 $\pm$ 200 K and the gravitational acceleration log$g$ = 8.17 $\pm$ 0.05 for HS0507+0434B. The ESO Supernova Ia Progenitor Survey (SPY) took high-resolution spectra of more than 1000 white dwarfs and pre-white dwarfs, including HS0507+0434B (Koester et al. 2009). The best atmospheric parameters of HS0507+0434B are $T_{eff}$ = 11488 $\pm$ 18 K and log$g$ = 8.057 $\pm$ 0.008. Largely based on the latest published version of the McCook \& Sion (1999) catalog for hydrogen-rich white dwarfs, Gianninas et al. (2011) made the spectral works over 1100 DA white dwarfs, including HS0507+0434B. For their best-fitting atmospheric model, $T_{eff}$ = 12290 $\pm$ 186 K and log$g$ = 8.24 $\pm$ 0.05. In order to compare the spectral results with each other conveniently, we show them in Table 4. The first two results are similar with each other while both them are smaller than the third results.

\section{Input physics and model calculations}

After reviewing the previous works on HS0507+0434B, we introduce our method in detail. The detailed period to period fittings need lots of white dwarf models in grids. White Dwarf Evolution Code (WDEC) is a "quasi-static" program, which computes a sequence of static models separated by finite steps in time. Each static model represents a white dwarf cooling at a different age and luminosity (Montgomery et al. 1999). WDEC can construct core composition profiles coupled with diffusion equilibrium helium shell and hydrogen atmosphere. We try to make white dwarf models with WDEC, taking the core composition profiles from MESA, which are results of nuclear burning process. WDEC makes the quasi-static evolutions from about 100000 K to the effective temperature we need. The program was firstly written by Schwarzschild and subsequently modified by Kutter \& Savedoff (1969), Lamb \& Van Horn (1975), and Wood (1990). For WDEC, the opacities are from Itoh et al. (1983). The equation of state (eos) is from Lamb (1974) in the degenerate and ionized core, and from Saumon et al. (1995) in the radiant and thin shell. There is a long time for a low-mass main-sequence star to evolve to the DAV stage. WDEC assumes the equilibrium profiles for H/He and He/C mixtures (Wood 1990). For the core compositions, WDEC can construct them by linear fitting to the real nuclear burning results. For MESA, the opacity tables are from Cassisi et al. (2007). The $\rho$-$T$ tables are based on Rogers \& Nayfonov (2002). In addition, MESA can make white dwarfs models of nuclear burning core compositions. The thermonuclear reaction rates are from Caughlan \& Fowler (1988, CF88) and Angulo et al. (1999, NACRE). Please see Paxton et al. (2011) for more details about MESA.

\begin{figure}
\begin{center}
\includegraphics[width=8cm,angle=0]{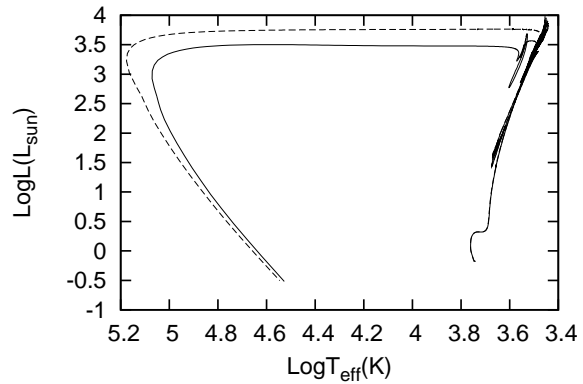}
\end{center}
\caption{MESA star evolution of 1.0 $M_{\odot}$. For the solid line, there are six-time thermal pulses. For the dashed line, there are fourteen-time thermal pulses.}
\end{figure}

\begin{figure}
\begin{center}
\includegraphics[width=8cm,angle=0]{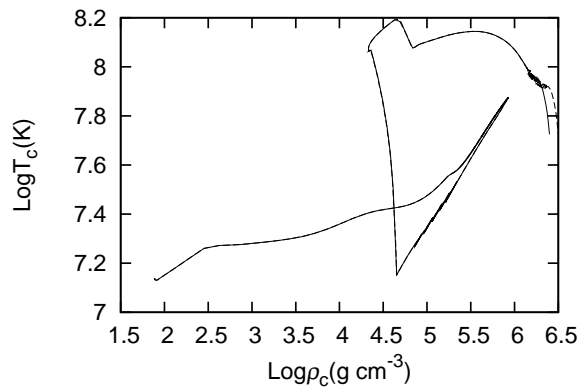}
\end{center}
\caption{Diagram of central density to central temperature.}
\end{figure}

\begin{figure}
\begin{center}
\includegraphics[width=8cm,angle=0]{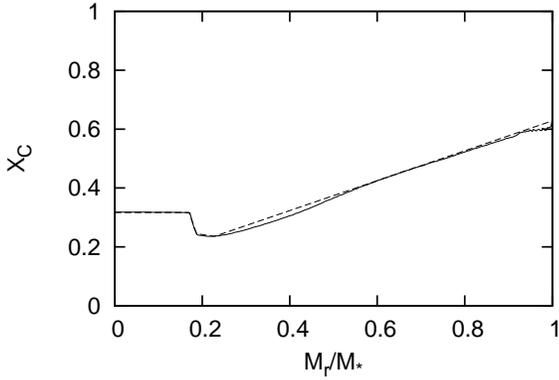}
\end{center}
\caption{Linear fitting to carbon profile of model0 from MESA.}
\end{figure}

\begin{figure}
\begin{center}
\includegraphics[width=8cm,angle=0]{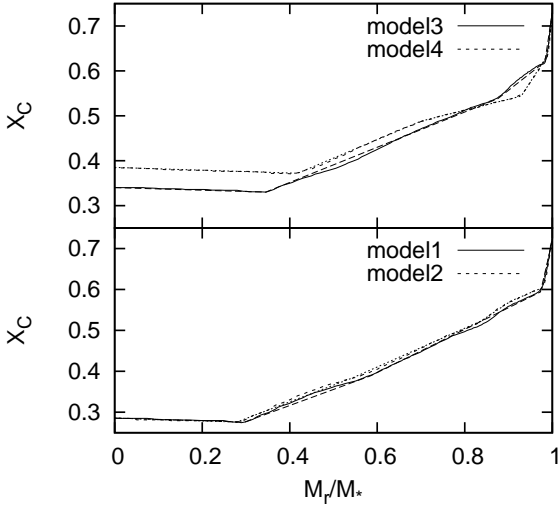}
\end{center}
\caption{Linear fittings to carbon profiles of model1, model2, model3, and model4 from MESA.}
\end{figure}

The version 4298 of MESA is downloaded and installed, which is a stable version after the test. We evolve one solar mass star from main-sequence to white dwarf stage (by using the file '1M\_pre\_ms\_to\_wd'). The initial parameters are 0.70 (hydrogen abundance), 0.02 (metal abundance), and 2.0 (mixing length parameter). The H-R diagram and the central density to central temperature diagram are shown in Fig. 1 and Fig. 2, respectively (solid lines). When the luminosity reaches -0.5, the evolution stops automatically. As shown in the document 'inlist\_1.0', the stellar wind is 'Reimers' for red giants (Reimers\_wind\_eta = 0.5), and 'Blocker' for AGB stars (Blocker\_wind\_eta = 0.05). The parameters are default settings. At the end of the evolution, the white dwarf model has log($M_{He}/M_{*}$) = -1.42, which is a little large. With LPCODE, Romero et al. (2012) made fully evolutionary white dwarf models. In order to decrease the helium content, they turned off the stellar wind during the thermal pulse stage and then they obtained thirty-time thermal pulses to obtain a thin helium layer. Adopting the similar method, we turn off the stellar wind during the thermal pulse stage and after fourteen-time thermal pulses, we try to reduce the helium layer mass. The H-R diagram and the central density to central temperature diagram are shown in Fig. 1 and Fig. 2, respectively (dashed lines). Then, log($M_{He}/M_{*}$) = -1.66, log($M_{H}/M_{*}$) = -3.90, and $M_{*}$ = 0.58 $M_{\odot}$. We try to fit the carbon profile of this white dwarf model. As shown in Fig. 3, the dashed lines are linear fittings to the solid curve, which is the carbon profile of the white dwarf model. The dashed lines are determined by four points (0.000, 0.317), (0.170, 0.317), (0.187, 0.245), and (1.000, 0.628). The ordinate is the carbon abundance and the abscissa is the mass fraction. In WDEC, the carbon profile can also be constructed by those four points and between them, there are straight lines. The oxygen abundance equals evidently to 1 - carbon abundance. Models with such a core composition are referred to as model0 in Table 5. In MESA, we can also make white dwarf models by using the file 'make\_co\_wd', with the stellar wind parameters (Reimers\_wind\_eta = 1 and Blocker\_wind\_eta = 5) being different from the file '1M\_pre\_ms\_to\_wd'. With the default settings and initial stellar mass of 2.0, 2.5, 3.0, and 3.5 $M_{\odot}$ respectively, we make white dwarf models (model1, model2, model3, and model4). The linear fittings to the carbon profiles of the four models are shown in Fig. 4 and the model information is displayed in Table 5.

The five models in Table 5 have different initial stellar masses, which have different nuclear burning histories and different core composition profiles. Model0 represents the evolution of a low-mass star, which has helium flash before the stable helium burning stage. The flat carbon profile in the central part of the core is a result of convective mixing. In addition, it requires a higher central temperature for a low-mass star than an intermediate-mass star to ignite helium because of the electron degeneracy. The neutrino energy losses cause that the highest temperature is beside the central region. This is the reason that the carbon has a decline process in Fig. 3. In Fig. 4, the carbon profiles look alike. As shown in the low panel of Fig. 4, model1 and model2 have nearly the same carbon profiles. Model2, model3, and model4 are evolution results of intermediate-mass stars. Though the initial mass changes from 2.0 $M_{\odot}$ to 3.5 $M_{\odot}$, the carbon profiles are similar on the white dwarf stage. A low-mass main-sequence star may evolve to be a white dwarf by strong stellar wind, while, an intermediate-mass star may evolve to be a white dwarf by the effect of binary mass exchanges. We try to compare these core compositions by doing asteroseismology works on HS0507+0434B.

\begin{table}
\begin{center}
\begin{tabular}{lllll}
\hline
Models         &Ms stars           &Wd stars             &log($M_{He}/M_{*}$) &log($M_{H}/M_{*}$) \\
\hline
model0         &1.0$M_{\odot}$     &0.584$M_{\odot}$     &-1.66               &-3.90              \\
model1         &2.0$M_{\odot}$     &0.580$M_{\odot}$     &-1.57               &-3.88              \\
model2         &2.5$M_{\odot}$     &0.593$M_{\odot}$     &-1.62               &-3.94              \\
model3         &3.0$M_{\odot}$     &0.633$M_{\odot}$     &-1.73               &-4.10              \\
model4         &3.5$M_{\odot}$     &0.704$M_{\odot}$     &-1.90               &-4.38              \\
\hline
\end{tabular}
\caption{Information of models.}
\end{center}
\end{table}

With linear fittings to the carbon profiles of the five models from MESA, we make white dwarf models in WDEC. The mixing length theory is from B\"{o}hm \& Cassinelli (1971) and Tassoul et al. (1990). The mixing length parameter, which is defined as a ratio of the mixing length to the pressure scale height, is adopted as 0.6. It is the same with which used by Bergeron et al. (1995). The mesh points in each stellar model are about 1000. When constructing white dwarf models, there are four adjustable parameters. A mass grid is from 0.600 $M_{\odot}$ to 0.800 $M_{\odot}$ with a step of 0.005 $M_{\odot}$. Log($M_{H}/M_{*}$) is from -10.0 to -4.0 with a step of 0.5. Log($M_{He}/M_{*}$) equals -4.0, -3.5, -3.0, -2.5, and -2.0, respectively. An effective temperature grid is from 11000 K to 12500K with a step of 50 K. With these grids, we make white dwarf models. And then, we solve the full equations of linear and adiabatic oscillation numerically. It finds each eigen-mode by scanning. With these methods and steps, we try to do the asteroseismology work on HS0507+0434B and discuss the effect of different core composition profiles.

\section{Our asteroseismology works on HS0507+0434B}

\begin{table}
\begin{center}
\begin{tabular}{lccccc}
\hline
                   &model0       &model1         &model2       &model3      &model4      \\
\hline
$T_{eff}$(K)       &11450        &11650          &11600        &12100       &11200       \\
log$g$             &8.088        &8.284          &8.284        &8.311       &8.289       \\
$M_{*}$/$M_{\odot}$&0.640        &0.780          &0.780        &0.780       &0.775       \\
log($q_{x}$)       &-6           &-4             &-4           &-10         &-5          \\
log($q_{y}$)       &-3           &-2             &-2           &-2.5        &-2          \\
$n\phi$(s)         &9.50         &14.97          &13.16        &10.50       &15.76       \\
$error_{big}$      &2.8          &8.1            &7.7          &4.7         &6.4         \\
                   &(553.7)      &(347.2)        &(347.6)      &(450.0)     &(662.3)     \\
\hline
\end{tabular}
\caption{The best-fitting models for different core compositions, $n\phi$s are included. $q_{x}$=$M_{H}$/$M_{*}$ and $q_{y}$=$M_{He}$/$M_{*}$.}
\end{center}
\end{table}

\begin{table}
\begin{center}
\begin{tabular}{llll}
\hline
$P_{obs}(s)$&$P_{mod}(s)$ & $\phi$(s) \\
\hline
748.6   &  750.2 & 1.583\\
697.6   &  699.7 &      \\
655.9   &  655.7 &      \\
556.5   &  553.7 &      \\
445.3   &  442.5 &      \\
355.3   &  355.3 &      \\
\hline
\end{tabular}
\caption{Our best-fitting model results, the core compositions of which are from model0 in Table 5.}
\end{center}
\end{table}

\begin{table}
\begin{center}
\begin{tabular}{llll}
\hline
$P_{obs}(s)$&$P_{mod}(s)$ & $\phi$(s) \\
\hline
748.6   &  747.6 & 1.750\\
697.6   &  695.9 &      \\
655.9   &  657.2 &      \\
556.5   &  556.2 &      \\
445.3   &  450.0 &      \\
355.3   &  356.8 &      \\
\hline
\end{tabular}
\caption{Our best-fitting model results, the core compositions of which are from model3 in Table 5.}
\end{center}
\end{table}

\begin{table}
\begin{center}
\begin{tabular}{lccccc}
\hline
                   &1.25      &1.50      &1.75      &2.75      &3.25      \\
\hline
$T_{eff}$(K)       &11900     &11800     &11750     &12350     &12400     \\
log$g$             &8.320     &8.320     &8.283     &8.296     &8.303     \\
$M_{*}$/$M_{\odot}$&0.785     &0.785     &0.780     &0.770     &0.775     \\
log($q_{x}$)       &-10       &-10       &-4        &-10       &-10       \\
log($q_{y}$)       &-2.5      &-2.5      &-2        &-2.5      &-2.5      \\
$n\phi$(s)         &14.41     &15.64     &11.84     &13.78     &12.32     \\
$error_{big}$      &5.0       &6.5       &9.1       &4.7       &5.2       \\
                   &(450.3)   &(451.8)   &(346.2)   &(450.0)   &(551.3)   \\
\hline
\end{tabular}
\caption{The best-fitting models for added different core compositions.}
\end{center}
\end{table}

\begin{table}
\begin{center}
\begin{tabular}{lccccc}
\hline
                     &Romero          &Fu        &Ours   \\
\hline
$T_{eff}$(K)         &12257$\pm$135   &12460     &11450  \\
log$g$               &8.10$\pm$0.06   &          &8.088  \\
$M_{*}$/$M_{\odot}$  &0.660$\pm$0.023 &0.675     &0.640  \\
log($q_{x}$)         &-4.43to-4.12    &-8.5      &-6     \\
log($q_{y}$)         &-1.92           &-2        &-3     \\
\hline
\end{tabular}
\caption{The best-fitting models for the asteroseismology works of Romero et al. (2012), Fu et al. (2013), and us. $q_{x}$=$M_{H}$/$M_{*}$ and $q_{y}$=$M_{He}$/$M_{*}$.}
\end{center}
\end{table}

\begin{figure}
\begin{center}
\includegraphics[width=8cm,angle=0]{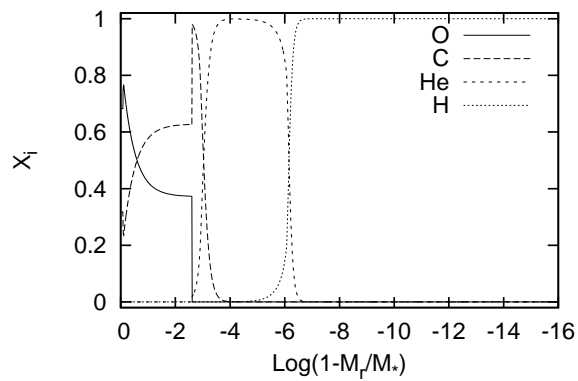}
\end{center}
\caption{The composition profiles of our best-fitting model.}
\end{figure}

\begin{table}
\begin{center}
\begin{tabular}{lllllll}
\hline
P(l,k)(s)    &  P(l,k)(s)    &P(l,k)(s)      &P(l,k)(s)     \\
\hline
195.98(1,1)  &1061.69(1,19)  &444.27(2,12)   &922.18(2,29)  \\
244.01(1,2)  &1123.29(1,20)  &465.46(2,13)   &958.18(2,30)  \\
284.46(1,3)  &1167.27(1,21)  &495.05(2,14)   &979.46(2,31)  \\
355.30(1,4)  &1219.08(1,22)  &530.94(2,15)   &988.37(2,32)  \\
404.02(1,5)  &1284.95(1,23)  &554.46(2,16)   &1021.62(2,33) \\
442.50(1,6)  &1325.07(1,24)  &584.46(2,17)   &1055.07(2,34) \\
507.61(1,7)  &1383.43(1,25)  &606.49(2,18)   &1081.36(2,35) \\
553.66(1,8)  &113.16(2,1)    &617.97(2,19)   &1119.78(2,36) \\
594.62(1,9)  &141.72(2,2)    &652.29(2,20)   &1148.56(2,37) \\
655.75(1,10) &192.86(2,3)    &676.15(2,21)   &1180.76(2,38) \\
699.67(1,11) &206.71(2,4)    &705.95(2,22)   &1214.94(2,39) \\
750.22(1,12) &233.62(2,5)    &743.57(2,23)   &1244.66(2,40) \\
792.48(1,13) &269.63(2,6)    &767.32(2,24)   &1281.78(2,41) \\
820.39(1,14) &295.56(2,7)    &801.73(2,25)   &1308.58(2,42) \\
859.57(1,15) &320.89(2,8)    &830.29(2,26)   &1342.92(2,43) \\
919.84(1,16) &353.43(2,9)    &862.38(2,27)   &1348.65(2,44) \\
960.87(1,17) &381.15(2,10)   &895.64(2,28)   &1380.47(2,45) \\
1014.43(1,18)&406.23(2,11)   &               &              \\
\hline
\end{tabular}
\caption{The pulsation periods of our best-fitting model.}
\end{center}
\end{table}

The parameter $\phi$ is very small (0.778 s) for the best-fitting model of Romero et al. (2012). But there are only four modes being fitted, including the suspicious mode of 743.40 s. With multi-site observations by Fu et al. (2013), there are six triplets. However, the resulted $\phi$ is 3.683 s for their best-fitting model, namely a total of 22.1 seconds error for the six modes. In addition, the white dwarf models of Fu et al. (2013) adopt the homogeneous core composition, namely C/O = 50/50. As mentioned above, we try to take the core profiles of linear fittings to the carbon profiles of model0, model1, model2, model3, and model4 from MESA. With WDEC, we also do the asteroseismology work on HS0507+0434B. Parameters of the five best-fitting models for different core compositions are shown in Table 6. Results of two best-fitting models (models of the smallest $n\phi$ and the second smallest $n\phi$) are shown in Table 7 and Table 8, the core compositions of which are from model0 and model3, respectively.

For the carbon profile of linear fitting to model0 from MESA, our best-fitting model parameters are $T_{eff}$ = 11450 K, log$g$ = 8.088, $M_{*}$ = 0.640 $M_{\odot}$, log($M_{H}$/$M_{*}$) = -6, and log($M_{He}$/$M_{*}$) = -3. The parameter $\phi$ is only 1.538 s, namely a total of 9.50 seconds error for the six modes. The effective temperature (11450 K) is in agreement with the atmospheric result of Fontaine et al. (2003) and Bergeron et al. (2004) ($T_{eff}$ = 11630 $\pm$ 200 K), and very close to the spectral result of Koester et al. (2009) ($T_{eff}$ = 11488 $\pm$ 18 K). The gravitational acceleration (log$g$ = 8.088) is very close to the result of Fontaine et al. (2003) and Bergeron et al. (2004) (log$g$ = 8.17 $\pm$ 0.05), also very close to the result of Koester et al. (2009) (log$g$ = 8.057 $\pm$ 0.008). For the carbon profile of linear fitting to model3 in Table 5, $\phi$ is 1.750 s for the best-fitting model, namely a total of 10.50 seconds error when fitting the six modes. Accordingly, the model parameters are $T_{eff}$ = 12100 K, log$g$ = 8.311, $M_{*}$ = 0.780 $M_{\odot}$, log($M_{H}$/$M_{*}$) = -10, and log($M_{He}$/$M_{*}$) = -2.5. The effective temperature and the gravitational acceleration are basically consistent with the spectral results of Gianninas et al. (2011), which use the Lemke (1997) Stark profiles.

In view of asteroseismology, $n\phi$ = 9.50 s is smaller than $n\phi$ = 10.50 s and they are in the same level. For the scenario of choosing model0, the fitting error reaches 2.8 s when using 553.7 s to fit 556.5 s and using 442.5 s to fit 445.3 s. For the scenario of choosing model3, the fitting error can reach 4.7 s when using 450.0 s to fit 445.3 s. If we choose model0, the mode of 286.1 s identified by Kotak et al. (2002) will be fitted by 284.46 s ($l$ = 1, $k$ = 3) and the mode of 743.40 s will be fitted by 743.57 s ($l$ = 2, $k$ = 23). If we choose model3, 286.1 s will be fitted by 271.05 s ($l$ = 1, $k$ = 3) and 743.40 s will be fitted by 751.03 s ($l$ = 2, $k$ = 23). In order to obtain a more clear conclusion, the progenitor mass grid is increased by 'make\_co\_wd'. First, a 2.90 $M_{\odot}$ main-sequence star is evolved to be a white dwarf. The carbon profile is very close to the carbon profile from progenitor 3.00 $M_{\odot}$ star. After all the grid-model calculations and detailed period to period fittings, we find the best-fitting model ($n\phi$ = 11.00 s) the same with the fifth column in Table 6. Then, we evolve a 3.10 $M_{\odot}$ main-sequence star to the white dwarf stage and fit the carbon profile to do the work. The carbon profiles are not so close that we obtain a different result. For the best-fitting model, $T_{eff}$ = 12500 K, log$g$ = 8.325, $M_{*}$ = 0.790 $M_{\odot}$, log($M_{H}$/$M_{*}$) = -8.5, log($M_{He}$/$M_{*}$) = -2.5, and $n\phi$ = 12.07 s. For model1 and model2, the carbon profiles are very close (the low panel in Fig. 4) and the best-fitting model parameters are basically the same. Therefore, we increase the progenitor mass grid in 1.25 $M_{\odot}$, 1.50 $M_{\odot}$, 1.75 $M_{\odot}$, 2.75 $M_{\odot}$, 3.25 $M_{\odot}$. The best-fitting model results are shown in Table 9. For progenitor main-sequence star of 1.25 $M_{\odot}$, 1.50 $M_{\odot}$, 1.75 $M_{\odot}$, 2.00 $M_{\odot}$, 2.50 $M_{\odot}$, and 3.50 $M_{\odot}$, the best-fitting models have effective temperatures about 11700 K. For progenitor main-sequence star of 2.75 $M_{\odot}$, 2.90 $M_{\odot}$, 3.00 $M_{\odot}$, 3.10 $M_{\odot}$, and 3.25 $M_{\odot}$, the best-fitting models have effective temperatures about 12300 K. For all the white dwarf models from 'make\_co\_wd', the best-fitting models have the total stellar masses about 0.780 $M_{\odot}$. For progenitor main-sequence star from 1.25 $M_{\odot}$ to 3.25 $M_{\odot}$, it is poor for the best-fitting models to fit 286.1 s and 743.40 s (in $l$ = 2 mode), like model3 mentioned above. For model4, there are 285.94 s ($l$ = 1, $k$ = 4) and 741.76 s ($l$ = 2, $k$ = 29). However, $n\phi$ = 15.76 for model4, which is a little larger than $n\phi$ = 9.50 s for model0. In the last row in Table 6 and Table 9, we show the big residual deviation and the poorly fitted mode for each scenario. Different modes have different residual deviations. It seems that the mode of 355.3 s (progenitor mass of 1.75 $M_{\odot}$, 2.00 $M_{\odot}$, and 2.50 $M_{\odot}$) and the mode of 445.3 s (progenitor mass of 1.25 $M_{\odot}$, 1.50 $M_{\odot}$, 2.75 $M_{\odot}$, and 3.00 $M_{\odot}$) are always poorly fitted. The effective temperature is commonly lower for the scenario of poor fitting 355.3 s (11750 K, 11650 K, and 11600 K) than the scenario of poor fitting 445.3 s (11900 K, 11800 K, 12350 K, and 12100 K). Therefore, the fitting results may be more sensitive to the effective temperature. If we increase the temperature grid near the best fitting model, the fitting results may be a little better but the modes of 286.1 s and 743.40 s will also be poorly fitted for these models. In all, the big residual deviation for model0 is the smallest (2.8 s).

All these asteroseismology results may indicate that choosing model0 to reflect the core composition of HS0507+0434B is more reasonable. Let's look at Table 6 and Table 9 again. The results clearly show that there is a significant impact on fitting results for different core compositions. Different core compositions can lead to very different asteroseismology results. Since model0 is better than the other models, we suggest that the DAV star HS0507+0434B is likely to come from a low-mass star, probably one solar mass star. The probability is relatively small for the binary mass exchanges. In addition, Gianninas et al. (2011) showed that HS0507+0434B was 48 pc from us and HS0507+0434A was 49 pc from us, which suggest that they are not physical double.

Then, we compare our best-fitting model to the previous asteroseismology results. The gravitational acceleration (log$g$ = 8.088) is in agreement with the result of Romero et al. (2012) (log$g$ = 8.10 $\pm$ 0.06). The total stellar mass ($M_{*}$ = 0.640 $M_{\odot}$) is also in agreement with the result of Romero et al. (2012) ($M_{*}$ = 0.660 $\pm$ 0.023 $M_{\odot}$). For the hydrogen layer mass and the helium layer mass, our best-fitting model shows log($M_{H}$/$M_{*}$) = -6 and log($M_{He}$/$M_{*}$) = -3. Results of Romero et al. (2012) showed $M_{H}$/$M_{*}$ = (5.68 $\pm$ 1.94)$\times$$10^{-5}$ (log($M_{H}$/$M_{*}$) = -4.43to-4.12) and $M_{He}$/$M_{*}$ = 1.21$\times$$10^{-2}$ (log($M_{He}$/$M_{*}$) = -1.92). Fu et al. (2013) gave log($M_{H}$/$M_{*}$) = -8.5 and log($M_{He}$/$M_{*}$) = -2. In addition, the effective temperature of our best-fitting model is also different from the previous asteroseismology works. we show the three asteroseismology results in Table 10.

The composition profiles of our best-fitting model are shown in Fig. 5. The carbon profile of the core is from Fig. 3, which is from model0 generated by MESA. The H/He and He/C mixtures are assumed to be diffusion equilibrium. The connection between the carbon in the core and the carbon in the interface of He/C is steep, which is an approximation. In addition, we show the detailed pulsation periods for $l$ = 1 and 2 modes of our best-fitting model in Table 11. We notice that there was also an identified single mode of low amplitude in Fu et al. (2013). It is 999.7 s, which was poorly fitted by 986.9 s ($l$ = 2, $k$ = 27) for the best-fitting model of Fu et al. (2013). It is also poorly fitted by 1014.43 s ($l$ = 1, $k$ = 18) or 988.37 s ($l$ = 2, $k$ = 32) for our best-fitting model. However, the 'linear combinations' of 197.7 s and 1382.7 s may also be pulsation mode frequencies, which were fitted by Fu et al. (2013) using $l$ = 2 modes. For our best-fitting model, 197.7 s may be fitted by 195.98 s ($l$ = 1, $k$ = 1) and 1382.7 s may be fitted by 1383.43 s ($l$ = 1, $k$ = 25), which are $l$ = 1 modes. The 'further signals' 703.9 s of high amplitude (9.06) were fitted by Fu et al. (2013) using 703.3 s ($l$ = 2, $k$ = 18). For our best-fitting model, the mode may be fitted by 705.95 s ($l$ = 2, $k$ = 22). Fu et al. (2013) tried to fit another 'further signals' 972.2 s of low amplitude using 972.8 s ($l$ = 2, $k$ = 26). For our best-fitting model, there is not a proper mode to fit 972.2 s. If it is really a pulsation mode frequency, the fitting error will be large when using the mode of 979.46 s ($l$ = 2, $k$ = 31) to fit it.

\section{Discussion and Conclusions}

Asteroseismology is used to detect the inner structure of stars, which is highly dependent on the input physics of stellar models. The core composition of a white dwarf is a result of nuclear burning process. So we decide to change the 'constructed carbon' profile into the 'real' nuclear burning profile. First, we evolve low-mass and intermediate-mass main-sequence stars to be white dwarfs in MESA. Then, we try to do linear fittings to carbon profiles of those models. At last, with the linear fitting core composition, added to the diffusion equilibrium helium shell and hydrogen atmosphere, WDEC can make white dwarf models in grids. With these grids of white dwarf models, we try to do the asteroseismology work on HS0507+0434B, which has been observed for many years. There was a total of six triplets for HS0507+0434B, which were identified as $l$ = 1 modes by rotation (Fu et al. 2013). We make white dwarf models from progenitor main-sequence stars of 1.00, 1.25, 1.50, 1.75, 2.00, 2.50, 2.75, 2.90, 3.00, 3.10, 3.25, and 3.50 solar mass. $n\phi$ = 9.50 s is the smallest, which is from 1.00 solar mass main-sequence star by '1M\_pre\_ms\_to\_wd'. Our best-fitting model parameters are $T_{eff}$ = 11450 K, log$g$ = 8.088, $M_{*}$ = 0.640 $M_{\odot}$, log($M_{H}$/$M_{*}$) = -6, and log($M_{He}$/$M_{*}$) = -3. There is a total of 9.50 seconds error for the six modes, which is smaller than 22.1 seconds error for the best-fitting model of Fu et al. (2013). In addition, our best-fitting model may also fit the identified mode of 286.1 s (Kotak et al. 2002) by 284.46 s ($l$ = 1, $k$ = 3) and the identified mode of 743.40 s (743.0 s) (Handler et al. 2002 (Kotak et al. 2002)) by 743.57 s ($l$ = 2, $k$ = 23).

For our best-fitting model, the effective temperature is in agreement with the atmospheric result of Fontaine et al. (2003) and Bergeron et al. (2004), and very close to the spectral result of Koester et al. (2009). Our gravitational acceleration is also very close to those atmospheric results and in agreement with the result of Romero et al. (2012). The total stellar mass is also consistent with the result of Romero et al. (2012). For the hydrogen layer mass, our best-fitting model shows the middle value between Romero et al. (2012) and Fu et al. (2013), as shown in Table 10. For the helium layer mass, our best-fitting model shows a thin value. We should judge the asteroseismology results combined with the spectral works and the fitting results. We choose the carbon profile of model0 as the best-fitting carbon profile, which means that HS0507+0434B is likely to come from a low-mass main-sequence star. During its evolution, there is likely to have no binary mass exchanges.

\section{Acknowledgements}
This work is supported by the Knowledge Innovation Key Program of the Chinese Academy of Sciences under Grant No. KJCX2-YW-T24 and Yunnan Natural Science Foundation (Y1YJ011001). We are very grateful to X. J. Lai, Q. S. Zhang, T. Wu and J. Su for their kindly discussion and suggestions.

\label{lastpage}

\end{document}